# Pulse propagation through a dispersive intracavity medium

H.N. Yum[1*], Y.J. Jang[1], and M.S. Shahriar[1,2]

[1]*Department of Electrical Engineering and Computer Science, Northwestern University, Evanston, IL 60208, USA*
[2]*Department of Physics and Astronomy, Northwestern University, Evanston, IL 60208, USA*
[*]*Corresponding author:* h-yum@northwestern.edu

**Abstract**: In this paper, we study theoretically the behavior of a pulse as it propagates through an intracavity fast-light medium. The method of using a transfer function to determine a pulse after it passes through a cavity is well known. However, this approach cannot be used to determine the behavior of the pulse *inside* the cavity. To circumvent this constraint, we use an approach that starts by finding a self-consistent solution for a monochromatic field of infinite spatial and temporal extents, and determine its amplitudes before, inside, and after the cavity. We then construct a Gaussian input pulse by adding a set of these waves, properly phased and weighted, to represent a moving pulse before the cavity. Adding these waves at various time intervals then yields the complete spatial profile everywhere, including before, inside and after the cavity. We first confirm the prediction of this model by analyzing the behavior of a pulse passing through an empty cavity, and comparing the prediction of the output with the one produced by the transfer function method. We then apply the technique to a cavity containing a fast-light medium. The resulting model allows us to visualize the behavior of the pulse as it propagates superluminally inside the cavity, and interferes with itself through multiple bounces. For a vanishing group index, an interference pattern is formed immediately after the pulse enters the cavity, with an output pulse emerging with no time delay or distortion. The results obtained here illustrates the physical mechanism behind pulse propagation through a white light cavity, a process we have proposed earlier for realizing a high bandwidth, long delay data buffering system.

It is well known that optical cavities can be used to control the group velocity of light. Various configurations of the cavities have been proposed to demonstrate slow-light or fast-light. For example, a Fabry Perot cavity has been investigated as a system for producing pulse delays[1]. Microresonators, such as coupled-resonator optical waveguides (CROWs)[2] and side-coupled integrated spaced sequence of resonators (SCISSOR)[3,4,5], have also been investigated for producing slow-light and fast-light. More recently, we proposed the use of so-called White Light Cavities (WLCs)[6,7,8,9], to realize a trap-door data buffer system where the delay time achievable far exceeds the limit imposed by the delay-bandwidth constraint encountered in a slow-light based data buffer[10,11].

Briefly, such a buffer consists of two WLCs in series, and works as follows. Each WLC is a high finesse cavity containing a medium inside which has a negative dispersion when a control beam is applied to it, or virtually no dispersion when the control beam is turned off. Initially, the control fields for both WLCs are turned off, so that the medium inside each WLC has virtually no dispersion, and the cavity transmission band is very narrow. A data sequence with a spectrum that is much broader than this band, and is shifted from the cavity resonance frequency by an amount larger than the pulse bandwidth, is thus fully reflected by the first cavity. When the control beam for the first WLC is turned on, its transmission band becomes wide enough to transmit the data sequence fully, without any distortion. After the data stream passes the first WLC, its control beam is turned off again. The data sequence now bounces back and forth between the two inactive cavities, undergoing very small attenuation in each pass. The data stream can be released by activating the control field for either WLC[10,11].

When the transmission band of the WLC becomes very broad, the group velocity in the intra-cavity medium is close to infinity over this bandwidth. The transfer function for the WLC which relates the output pulse to the input pulse can be readily determined analytically[10]. However, the behavior of the pulse as it propagates superluminally while *inside* the cavity cannot be determined using the transfer function approach. In this paper, we develop a new approach to visualize the behavior of the pulse inside any cavity, and apply this approach to interpret the propagation of pulses through a WLC. In this approach, we start by finding a self-consistent solution for a monochromatic field of infinite spatial and temporal extents, and determine its amplitudes before, inside, and after a cavity. We then construct a Gaussian input pulse by adding a set of these waves, properly phased and weighted, to represent a moving pulse before the cavity. Adding these waves at various time intervals then yields the complete spatial profile everywhere, including before, inside and after the cavity. In particular, it reveals waves in both forward and backward directions, including multiple bounces occurring inside the cavity. This approach is generic, and can be applied to any situation, including an empty cavity. We first confirm the prediction of this model by analyzing the behavior of a pulse passing through an empty cavity, and comparing the prediction of the output with the one produced by the transfer function method. We then apply the technique to the WLC. The output pulse produced this way again is found to agree with the prediction of the transfer function method. The resulting model allows us to visualize the behavior of the pulse as it propagates superluminally inside the WLC.

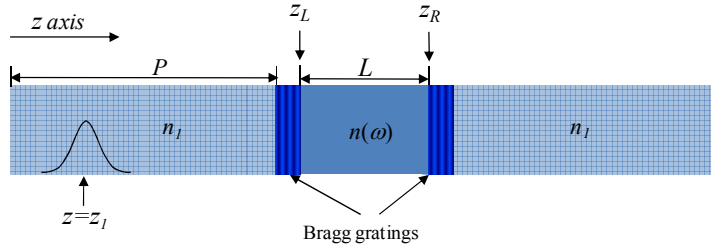

Fig.1.Schematic illustration of a typical Fabry-Perot (FP) cavity of length L, containing a dispersive medium. We model the mirrors as Bragg gratings. The medium outside the FP is assumed to be non-dispersive. At $t=t_1$, the peak of a test pulse is located at $z=z_1$, moving in the positive z direction.

The generic system we consider here is illustrated schematically in Fig.1. A dispersive medium of length $L$ is placed between two mirrors. For concreteness[12], we model each mirror to be a Bragg grating (BG), each with an intensity reflectivity $R$ that is assumed to be constant over the bandwidth of a test pulse. The medium outside the cavity, on each side, is assumed to be non-dispersive, and of infinite extent. The origin of the z-coordinate is set to be at a distance $P$ from the left BG. The mean indices of the dispersive medium and the non-dispersive medium are $n_0$ and $n_1$, respectively. When $t=t_1$, a Gaussian pulse with a finite spatial extent is centered at $z=z_1$, and moves in the positive z direction. We decompose the pulse into monochromatic waves, calculate phase changes during the propagation through the FP cavity, and use inverse Fourier transform to find the resultant pulse envelope in z domain. To this end, we consider first a monochromatic input wave of *unity* amplitude, expressed as:

$$E_{in}(\omega,z,t) = exp(j\phi_{in}) \text{ where } \phi_{in} = k_1(z-z_1) - \omega(t-t_1) \quad (1)$$

where $\omega$ is the angular frequency, and $k_1 = n_1\omega/c$, where $c$ is the speed of light in vacuum. In this notation, $\phi_{in}$ represents the net phase change at an arbitrary spatio-temporal coordinates $\{z, t\}$, relative to the reference coordinates $\{z_1, t_1\}$.

The wave is reflected by the BG on the left at $z=P$, and then propagates backward. At an arbitrary spatial point $z$, the reflected wave has accumulated a net spatial phase of $k_l(P-z_l)+k_l(P-z)$. In addition, note that, according to the coupled wave theory[13,14], if a wave is reflected by a BG with an arbitrary Bragg-reflection coefficient, then the reflected wave is shifted in phase by $-\pi/2$.[15] Furthermore, we assume that the BG thickness is infinitesimally small, leading to a Bragg spectral width much larger than that of the FP, so that the reflectivity, $\sqrt{R}$, can be assumed to be uniform over the bandwidth of interest[15]. The reflected wave therefore can be written as:

$$E_{rl}(\omega,z,t) = \sqrt{R}\exp\left(-\frac{j\pi}{2}\right)\exp(j\phi_{rl}); \quad \phi_{rl} = k_l(2P-z-z_l)-\omega(t-t_l) \tag{2}$$

Note that $\phi_{rl}(z=P) = \phi_{in}(z=P)$.

A part of the wave is also transmitted through the left BG and propagates inside the FP cavity. It is reflected again by the right BG, and returns to the left BG. Therefore, we have to consider two sets of counter-propagating. The forward-propagating waves inside the cavity are given by the following infinite series summation:

$$E_f(\omega,z,t) = \sum_{m=0}^{\infty}\left[R^m \exp(-jm\pi)\exp(2jmk_d L)\sqrt{T}\exp(j\phi_f)\right]; \quad \phi_f = k_l(P-z_l)+k_d(z-P)-\omega(t-t_l) \tag{3}$$

where $T$ is the transmittance ($R+T=1$), and $k_d$ is the wave number in the dispersive medium so that $k_d = n(\omega)\omega/c$. Note that $\phi_f(z=P) = \phi_{in}(z=P)$.

To produce anomalous dispersion, we consider dual gain peaks centered around $\omega_0$, the empty-cavity resonance frequency. The anti-symmetric profile for $n(\omega)$ allows us to set $n_{02} \equiv (1/2)d^2n/d\omega^2\big|_{\omega=\omega_0} = 0$. Thus, $n(\omega)$ is expressed in terms of a Taylor expansion around $\omega_0$: $n(\omega) = n_0 + (\omega-\omega_0)n_{01} + (\omega-\omega_0)^3 n_{03}$, where $n_{01} = dn/d\omega\big|_{\omega=\omega_0}$, and $n_{03} = (1/6)dn^3/d\omega^3\big|_{\omega=\omega_0}$. $\phi_f$ contains the phase terms: $k_l(P-z_l)$ in the lhs non-dispersive medium and $k_d(z-P)$ in the intracavity dispersive medium. Similarly, the backward-propagating waves inside the cavity can be expressed as:

$$E_b(\omega,z,t) = \sum_{m=0}^{\infty}\left[R^{m+\frac{1}{2}}\exp\left(\frac{-j(2m+1)}{2}\pi\right)\exp(2jmk_d L)\sqrt{T}\exp(j\phi_b)\right]; \quad \phi_b = k_l(P-z_l)+k_d L+k_d(P+L-z)-\omega(t-t_l)$$

$$\tag{4}$$

After every bounce inside the cavity, the wave is transmitted through the left BG and then propagates in the $-\hat{z}$ direction. These beams, added together, produce the net additional reflected beam given by:

$$E_{r2}(\omega,z,t) = \sum_{m=0}^{\infty}\left[R^{m+\frac{1}{2}}\exp\left(\frac{-j(2m+1)}{2}\pi\right)\exp(2jmk_dL)T\exp(j\phi_{r2})\right]; \quad \phi_{r2} = k_1(P-z_1)+2k_dL+k_1(P-z)-\omega(t-t_1) \quad (5)$$

Note that the total reflected field is given by the sum of $E_{r1}$ and $E_{r2}$. Note also that the $E_{r2}(z=P)=\sqrt{T}E_b(z=P)$, and $\phi_{r2}(z=P)=\phi_b(z=P)$. The difference between $\phi_{r2}$ and $\phi_b$ takes into account the additional phase accumulated via propagation in the non-dispersive medium with wave number $k_1$.

Similarly, using Eq.(3), the overall transmitted field outside the right BG can be expressed as:

$$E_{out}(\omega,z,t) = \sum_{m=0}^{\infty}\left[R^m\exp(-jm\pi)\exp(2jmk_dL)T\exp(j\phi_{out})\right]; \quad \phi_{out} = k_1(P-z_1)+k_dL+k_1(z-P-L)-\omega(t-t_1) \quad (6)$$

Note that $E_{out}(z=P+L)=\sqrt{T}E_f(z=P+L)$, and $\phi_{out}(z=P+L)=\phi_f(z=P+L)$. As before, the difference between $\phi_{out}$ and $\phi_f$ takes into account the additional phase accumulated via propagation in the non-dispersive medium with wave number $k_1$. Finally, note that the phase terms resulting from time evolution in Eqs.(1) through (6) have the same sign (minus) to assure that the phase velocity is $+c$ for $E_{in}$, $E_f$ and $E_{out}$, and $-c$ for $E_{r1}$, $E_b$ and $E_{r2}$.

Summing the infinite series in the expressions above, we find:

$$E_f(\omega,z,t) = \frac{\sqrt{T}}{1-R\exp(-j\pi)\exp(2jk_dL)}\exp(j\phi_f) \quad (7.a)$$

$$E_b(\omega,z,t) = \frac{\exp(-j\pi/2)\sqrt{R}\sqrt{T}}{1-R\exp(-j\pi)\exp(2jk_dL)}\exp(j\phi_b) \quad (7.b)$$

$$E_{r2}(\omega,z,t) = \frac{\exp(-j\pi/2)\sqrt{R}T}{1-R\exp(-j\pi)\exp(2jk_dL)}\exp(j\phi_{r2}) \quad (7.c)$$

$$E_{out}(\omega,z,t) = \frac{T}{1-R\exp(-j\pi)\exp(2jk_dL)}\exp(j\phi_{out}) \quad (7.d)$$

Next we consider a Gaussian pulse, centered at $z=z_1$ at $t=t_1$, with a spatial width of $z_0$. The amplitude of the spatial frequency spectrum of this pulse can be expressed as $\tilde{S}_0(k)=\left(z_0/\sqrt{2}\right)\exp\left[-(k-k_0)^2z_0^2/4\right]$ where $k_0=\omega_0/c$,

$k = \omega/c$. Each monochromatic component of this pulse acts as an input field of the type expressed in Eq. 1. Applying inverse Fourier transform (IFT), and using $dk = d\omega/c$, we can now find the pulse everywhere in $z$-space:

$$\text{For } 0 < z < P, \; S_1(z,t) = 1/\sqrt{2\pi} \int_{-\infty}^{\infty} \tilde{S}_0(k) \left[ \sum_{i=in,r1,r2} E_i(\omega,z,t) \right] dk \quad (8.a)$$

$$\text{For } P < z < P+L, \; S_2(z,t) = 1/\sqrt{2\pi} \int_{-\infty}^{\infty} \tilde{S}_0(k) \left[ \sum_{i=f,b} E_i(\omega,z,t) \right] dk \quad (8.b)$$

$$\text{For } z > P+L, \; S_3(z,t) = 1/\sqrt{2\pi} \int_{-\infty}^{\infty} \tilde{S}_0(k) E_{out}(\omega,z,t) dk \quad (8.c)$$

In what follows, we assume, without loss of generality, a situation where $n_0 = n_1 = 1$. First, it is instructive to discuss the pulse propagation in a non-dispersive intracavity medium by setting $n_g = 1$. Fig.2 illustrates the pulse propagating along the $z$ axis, as freeze-frames at different times. For illustration, we consider $R$=0.7, $T$=0.3, $P$=500, $L$=93, $z_1$=200, $z_0$=10.3746, free spectral range(FSR)=1.613MHz, and $\omega_0$=2π×1.93×10$^{14}$ s$^{-1}$ corresponding to the wavelength λ of 1550nm, where all distances are in meters. To find the resonance condition for the cavity, it is important to consider the additional phase term due to the Bragg reflections: $exp(-j\pi)$ in the denominator of Eq. (7d). Thus, the cavity resonates on the condition that $L = c(2q+1)\pi/(2n_0\omega_0)$ where q is a positive integer. To observe graphically the pulse inside the cavity, q is chosen to fulfill the condition that $L > z_0$.

In Fig. 2(a), the incoming pulse is shown at $t = 250/c$. Fig. 2(b) illustrates that for $t = 300/c$, the pulse is reflected from the left BG. Note that the interference between $E_{in}$ and $E_{r1}$ is seen in $0 < z < P$. In Fig.2(c), the incident pulse is split into two pulses after hitting the left BG: the wave given by $E_{r1}$ produces the reflected pulse in $0 < z < P$, and the wave given by $E_f$ produce the forward-propagating pulse inside the cavity. For $t = 390/c$, $E_f$ is reflected by the right BG and then propagates backward, resulting in the backward propagating intra-cavity field $E_b$. The waves given by $E_b$ and $E_f$ interfere around the front of the right BG, as displayed in Fig. 2(d). These interference patterns are on the scale of the wavelength of light, while the length scales we are considering are macroscopic; as such, the true pattern is masked by the coarseness of sampling. In order to show the pattern clearly, we plot the expanded views over $z_2 < z < z_2 + \lambda$ and $z_3 < z < z_3 + \lambda$ over a distance of one wavelength ($\lambda = 1.55\mu$m) in the insets of Fig.2(b) and (d), respectively, where $z_2 = z_L - \lambda$ ($z = 500 \equiv z_L$) is a position one wavelength ahead of the left end of the cavity, and

$z_3 = 590$ is a position that is more than 3 meters ahead of the right end of the cavity. These parameters have no particular significance, and have been chosen for illustrative convenience only. Fig. 2(e) illustrates that for $t = 650/c$ the backward-propagating pulse ($E_b$) within the cavity moves toward the left BG. The two pulses in sequence associated with $E_{out}$ propagate with the separation equal to $2L$, which is the optical path resulting from a round trip inside the cavity. The leading pulse in the output pulse train is observed with no delay compared to the free-space propagating reference (blue) and with the intensity attenuation of 91% due to the cavity reflection.

Note that in this simulation we have used a pulse bandwidth ($\Delta v_{pulse}$) that is a factor of 50 larger than the bandwidth of the cavity ($\Delta v_{FWHM}$). Since the overlap between the pulse spectrum with the cavity tranmission spectrum is negligible, the pulse essentially undergoes sequential reflections at the two BGs, without any significant multi-beam interference. As a result, we do not observe any noticeable distortion, nor delay. In order to verify this results, we have also determined the output pulse using a transfer function method[1,10,11], in time domain, as shown in figure 2(f). Here, we see two output pulses in series, with a temporal separation of $2L/c$, in agreement with the distance $2L$ between the two output profiles in the spatial domain, shown in Fig. 2(e).

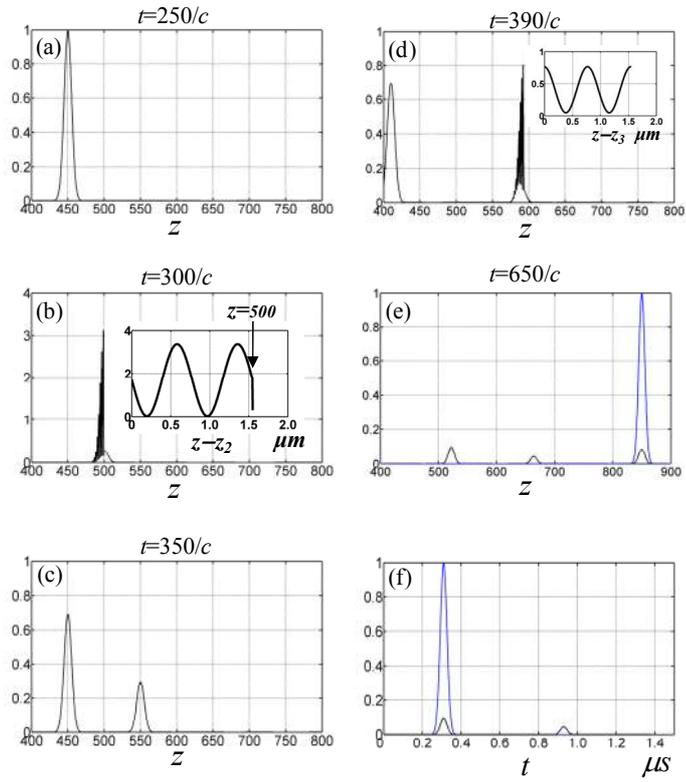

Fig.2. A propagating pulse shown as freeze-frames for (a) $t = 250/c$, (b) $t = 300/c$, (c) $t = 350/c$, (d) $t = 390/c$, (e) $t = 650/c$. (f) Numerical simulation by a transfer function method. The insets present expanded views of the interference patterns. In Fig. (e) and (f), the reference pulse (blue) propagates together with the first output from the cavity. Note that the left BG is at $z = 500 \equiv z_L$ and the right BG is at $z = 593 \equiv z_R$.

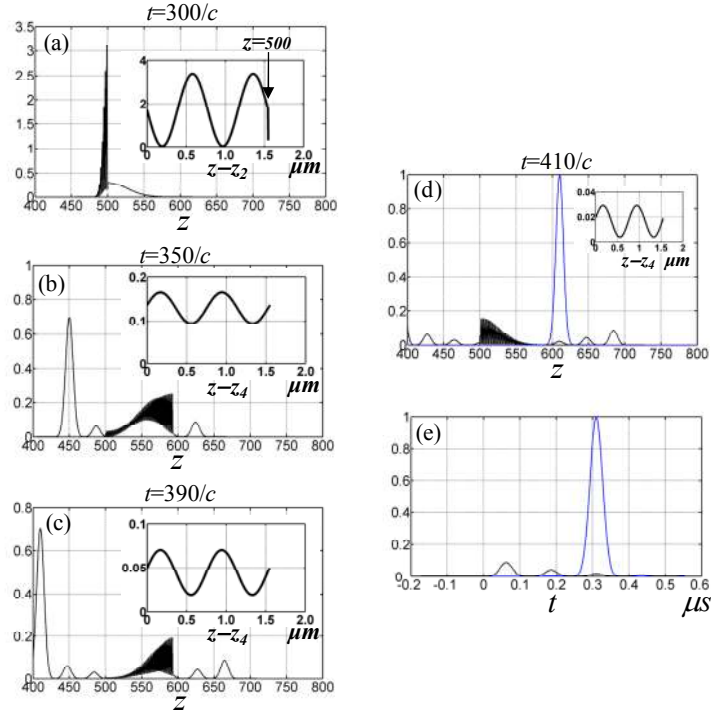

Fig. 3. Propagation of a pulse shown as freeze-frames, for an intracavity fast-light medium with group index $n_g = 0.2$, for (a) $t = 300/c$, (b) $t = 350/c$, (c) $t = 390/c$, (d) $t = 410/c$ with a reference pulse(blue). The insets present expnaded views over one wavelength. (e) Output pulses and the reference in time domain.

We consider next the behavior of a pulse propagating through a fast-light intracavity medium, for various sub-unity values of the group index, $n_g$. We employ the same cavity parameters as those used in Fig. 2, and consider only the first order dispersion ($n_{01}$) in the Taylor expansion of $n(\omega)$. Fig. 3 illustrates the propagation of a pulse for $n_g = 0.2$. At $t = 300/c$, the value of $n_g$ determines the distance between a front edge of the pulse and the right BG. The interference pattern over $z_2 < z < z_2 + \lambda$ are expanded in the inset of Fig. 3(a). The insets of Fig.3(b)~(d) illustrate the intracavity interference over $z_4 < z < z_4 + \lambda$, where $z_4 = 550$ represent a position near the center of the cavity. Comparing Fig. 3(a) with Fig. 2(b), both corresponding to $t = 300/c$, we can see the fast-light effect manifestly. In Fig. 2(b), there is no fast-light effect, and the pulse is not stretched. In Fig. 3(a), the group velocity is superluminal, and the front end of the pulse has almost reached the right BG. The intracavity fast-light medium also causes the counter propagating pulses to expand. Thus, the forward-propagating pulse associated with $E_f$ and the backward-propagating pulse associated with $E_b$ interfere with each other over the entire spatial range inside the cavity, as illustrated in Fig.3(b)~(d). Fig. 3(b) and (c) illustrate that the pulse train is composed of the reflected pulse associated with $E_{r1}$ (the first pulse on the left of the cavity), a series of pulses resulting from $E_{r2}$ (remaining pulses on the left of the cavity) and $E_{out}$ on the right of the cavity. Of course, the separation in *z*-axis between the adjacent pulses corresponds to the optical path for one round trip inside the cavity i.e. $2n_g L$. Fig. 3(d) shows that on the right side of the cavity ($z > 593$) the first and the second pulses are advanced by $(n_0 - n_g)L$ and $(n_0 - 3n_g)L$, respectively, compared to the free-space propagating reference, and the third pulse coincides with the reference, since $(n_0 - 5n_g)L = 0$. Note that $(2m+1)n_g L$ is the optical path in the intracavity medium where $m$ is the number of round trips after the pulse hits the right BG for the first time. In Fig. 3(e), a transfer function method is used to produce the output pulse train in time domain. The pulses appear with a repetition rate of $2n_g L/c$, in agreement with the spatial profiles shown in figure 3(d). By comparison with the reference, we find that the time advancement also agrees with the results displayed in Fig. 3(d). For example, the delay time between the first pulse and the reference is $(n_0 - n_g)L/c$.

Fig. 4 illustrates the cases of $n_g \simeq 0$ corresponding to the ideal WLC condition. In Figs. 4(a)~(c), the portion of the pulse that is loaded into the intracavity fast-light medium passes instantly through the cavity. Subsequently, it shows up to the right side after the cavity (z>593) before the rest of the pulse enters into the cavity. Of course, such a fast movement inside the cavity is the obvious manifestation of $v_g \simeq \infty$.[16]

In order to understand the behavior of the pulse in this case, it is instructive first to consider the limit where the cavity mirrors are eliminated. In the simulation, this can be achieved simply by setting *T*=1, *R*=0 for each BG, corresponding to a free-space zone of infinite group velocity (ZIGV) from z=500 ($\equiv z_L$) to z=593 ($\equiv z_R$). The resulting profiles are shown in the upper insets of each Figs. 4(a)~(c). As can be seen, when the pulse enters the ZIGV, it gets split immediately, with the front part of the pulse appearing outside this zone on the right. The amplitude of the field in the ZIGV becomes a constant, equaling the value of the pulse amplitude at the point of splitting. More explicitly, we denote by $f(z)$ the shape of the pulse, for $n_g = n_0$, at a time when it has crossed the point $z = z_L$. Then, for $n_g = 0$, we have:

$$\begin{aligned} z_L \geq z: &\quad E(z) = f(z) \\ z_L < z < z_R: &\quad E(z) = f(z_L) \\ z_R \leq z: &\quad E(z) = f(z - (z_R - z_L)) \end{aligned} \quad (9)$$

The physical meaning of the flat profile inside the ZIGV can be understood by considering the fact that $d\phi/d\omega \equiv (\ell/c)d(n(\omega)\omega)/d\omega = n_g \ell/c$, where $\phi$ is the phase shift due to the propagation through the medium. For $n_g = 0$, we thus have $d\phi/d\omega = 0$, so that $\phi$ is constant. The value of the pulse envelope, for example ~0.2 at $z = z_L$ as shown in the upper inset of Fig. 4(a), is determined by summing the constituent waves. If the phase differences among the waves before the ZIGV are preserved, the pulse envelope value then does not change during the propagation through the ZIGV. As a result, the envelope value exhibits the height equal to that at $z = z_L$, for the whole ZIGV, as illustrated in the inset of Fig. 4(a).

This result can also be seen formally from Eqn. (7a), by setting $R=0$, $T=1$. We then get $E_f = exp(j\phi_f)$, where $\phi_f$ is the phase shift for each wave as it traverses the dispersive zone. For $n_g = 0$, this phase shift is the same for each constituent wave, independent of frequency, as noted above. For the case of the cavity (i.e., $R \neq 0, R + T = 1$), the resonance condition [$2k_d L = (2m+1)\pi$] is satisfied for each constituent wave, independent of frequency, for the ideal WLC condition of $n_g = 0$. This can be seen by noting that $k_d = n(\omega)\omega/c$ that, so that $\partial k_d / \partial \omega = (1/c)\partial[n(\omega)\omega]/\partial \omega = (1/c)n_g = 0$, which is akin to the fact that when the group velocity becomes infinite, the wavelength (and therefore the wave number) becomes independent of frequency. As a result, we find from Eqn. (7a) that $E_f = 1/\sqrt{T} \, exp(j\phi_f)$, with the value of $\phi_f$ being the same for each constituent wave. Thus, the pulse amplitude for the forward propagating pulse inside the cavity is also a constant, but larger by a factor of $1/\sqrt{T}$ when compared to the case of free space propagation through the ZIGV. This enhancement is the same as what would be seen for a continuous wave passing through such a cavity on resonance, which results from constructive interference between the infinite number of bounces it undergoes. Here, the forward wave is moving at an effectively infinite velocity, thus carrying out the infinite number of bounces immediately, and yielding the amplitude that would result from the constructive interference of these bounces, since each frequency component is resonant.

Consider next the backward-propagating wave, $E_b$, as expressed in Eq. (7b) for each constituent wave. Again, for $n_g = 0$, each wave is resonant, independent of frequency, and the propagation phase, $\phi_b$, is the same for each wave, independent of frequency. Thus, within the ZIGV (i.e., within the cavity in this case), the reflected wave is also a constant. For each bounce, the backward-propagating wave is smaller than the corresponding forward propagating wave by a factor of $\sqrt{R}$, due to the reflection from the right BG, and has relative, constant phase factor of $exp(-j\pi/2)$ due to this reflection. However, just as in the case of the forward propagating waves, the backward propagating waves also undergo constructive interference between the infinite number of reflections immediately, forming a net wave which is smaller than the net forward wave, $E_f$, by a factor of $\sqrt{R}$, and differed by a constant phase factor of $exp(-j\pi/2)$. Thus, the net field inside the cavity is the result of the interference between these two waves of constant amplitudes, producing a standing wave and a traveling wave (since the amplitudes are different). Of course, the amplitudes of these standing and traveling wave components change with time as the input pulse keeps moving past the entrance at $z = z_L$, as can be seen in Figs. 4(a) ~(c). The intracavity net field can thus be expressed in general as: $f(z_L)\left[1/\sqrt{T}\,exp(j\phi_f) + \sqrt{R}/\sqrt{T}\,exp(j\phi_b)\right]$. We plot in the lower insets the expanded views over $z_4 < z < z_4 + \lambda$ where again $z_4 = 550$ is a point near the center of the cavity. Note that the maximum ($\equiv E_{max}$) of the oscillating field amplitude corresponds to $f(z_L)\left[1/\sqrt{T} + \sqrt{R}/\sqrt{T}\right]$ and the minimum ($\equiv E_{min}$) to $f(z_L)\left[1/\sqrt{T} - \sqrt{R}/\sqrt{T}\right]$. In particular, the lower inset in Fig. 4(b) indicates that for $f(z_L) \simeq 1$, $|S_2(z,t)|^2$ oscillates between $|E_{min}|^2 = 0.089$ and $|E_{max}|^2 = 11.24$.

Another important aspect of the result shown in Fig. 4 is the absence of any reflection by the cavity. Of course, this follows from the fact that each constituent wave is resonant for the ideal WLC condition of $n_g = 0$, as discussed above. We can also see this explicitly by considering the expressions for $E_{r1}$ given by Eq. 2 and $E_{r2}$ given by Eq. 7(a), keeping in mind that the net reflected beam is given by the sum of these two. By setting $2k_dL = (2m+1)\pi$ on resonance in Eq.(7c), we get $E_{r2} = exp(-j\pi/2)\sqrt{R}exp(j\phi_{r2})$ where $\phi_{r2} = k_1(P-z_1) + k_1(P-z) - \omega(t-t_1) + \pi$. Comparing this to Eqn. (2), we see that $E_{r2} = -E_{r1}$, so that the net reflected field is zero. Since this conclusion is true for each constituent wave, there is no reflection by the ideal WLC.

Finally, Fig. 4(d) illustrates that the pulse emerging from the ideal WLC (black trace) is advanced compared to a reference pulse (blue trace) propagating through a non-dispersive medium of index $n_0$. The distance of advancement $(n_0 - n_g)L = n_0L$, as expected, meaning that the pulse suffers no time delay at all in crossing the WLC. In Fig. 5(e), we show the corresponding result in time domain, again calculated by using the transfer function method. The advance in time compared to the reference pulse is in agreement with the result of Fig. 4(d). Of course, we would see the same amount of advancement, in time or space, if we did not use a cavity, but still used a dispersive medium with $n_g = 0$. As noted in the introduction, the use of the cavity lies in the fact that if the anomalous dispersion is turned off, a pulse with the center of its spectrum shifted from the resulting narrow-band cavity with vanishing overlap would be fully reflected, making the process useful for trap-door data buffering[10]. Finally, we note that this type of advancement using a free space medium cannot be used to violate causality for any real system[17,18]; the same conclusion holds for an ideal WLC.

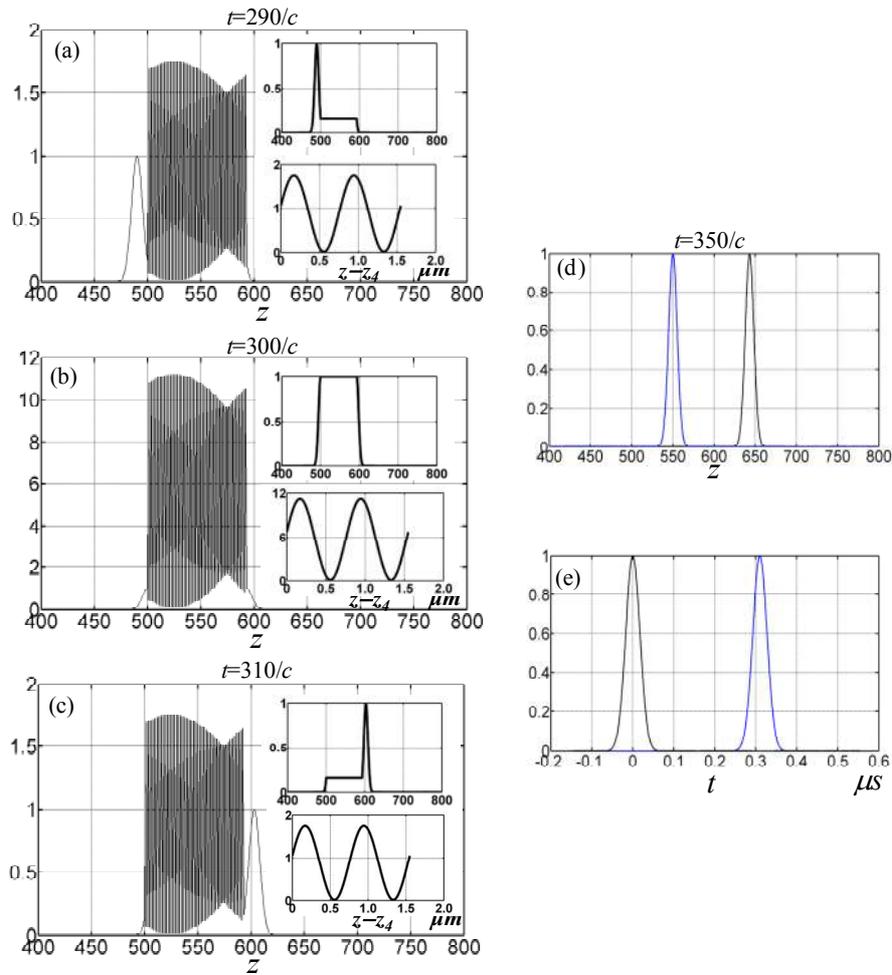

Fig.4. Propagation of a pulse shwon as freeze-frames, through an intracavity fast-light medium under ideal WLC condition ($n_g = 0$) for (a) $t = 290/c$, (b) $t = 300/c$, (c) $t = 310/c$, (d) $t = 350/c$, The upper insets illustrate the case of a fast-light medium with $n_g = 0$ in free space. The lower insets show the views expanded horizontally close to $z_4$=550. (e) Numerical simulations for cavity output (black) and the reference (blue) in time domain, produced via the transfer function method.

To summarize, in order to understand the behavior of a pulse inside a cavity, we have developed an approach that starts by finding a self-consistent solution for a monochromatic field of infinite spatial and temporal extents, and determine its amplitudes before, inside, and after the cavity. We then construct a Gaussian input pulse by adding a set of these waves, properly phased and weighted, to represent a moving pulse before the cavity. Adding these waves at various time intervals then yields the complete spatial profile everywhere, including before, inside and after the cavity. In particular, it reveals waves in both forward and backward directions, including multiple bounces occurring inside the cavity. This approach is generic, and can be applied to any situation, including an empty cavity. We first confirm the prediction of this model by analyzing the behavior of a pulse passing through an empty cavity, and comparing the prediction of the output with the one produced by the transfer function method. We then apply the technique to a cavity containing a fast-light medium. The output pulse produced this way again agrees with the prediction of the transfer function method. The resulting model allows us to visualize the behavior of the pulse as it propagates superluminally inside the cavity, and interferes with itself through multiple bounces. For the limiting case of a vanishing group index over the entire bandwidth of the pulse, an interference pattern is formed immediately after the pulse enters the cavity, with an output pulse emerging with no time delay or distortion. The results obtained here illustrates the physical mechanism behind pulse propagation through a white light cavity, a process we have proposed earlier for realizing a high bandwidth, long delay data buffering system.

[1] J. Yu, S. Yuan, J.Y. Gao and L. Sun, "Optical pulse propagation in a Fabry-Perot etalon: analytical discusstion," J. Opt. Soc. Am. A **18** 2153-2160 (2001)

[2] J.K.S Poon, J. Scheuer, S. Mookherjea, G.T. Paloczi, Y. Huang, and A. Yariv, "Matrix analysis of microring coupled-resonator optical waveguides," Opt. Exp. **12** 90-103 (2004)

[3] J.E. Heebner, R.W. Boyd and Q.H. Park, "SCISSOR solitons and other novel propagation effects in microesonantor-modified waveguides," J. Opt. Soc. Am. B **19** 722-731 (2002)

[4] J.E. Heebner and R.W. Boyd, " 'Slow' and 'fast' light in resonator-coupled waveguides," J. Mod. Opt. **49** 2629-2636 (2002)

[5] J.E. Heebner, R.W. Boyd and Q.H. Park, "Slow light, induced dispersion, enhanced nonlinearity, and optical solitons in a resonantor-array waveguides," Phys. Rev. E **65** 036619

[6] G.S. Pati, M. Messal, K. Salit, M.S. Shahriar, "Demonstration of a Tunable-Bandwidth White Light Interferometer using Anomalous Dispersion in Atomic Vapor," Phys. Rev. Lett. **99**, 133601 (2007).

[7] A. Wicht, K. Danzmann, M. Fleischhauer, M. Scully, G. Miller, and R. H. Rinkleff, "White-light cavities, atomic phase coherence, and gravitational wave detectors," Opt. Commun. **134**(1-6), 431–439 (1997).

[8] R. H. Rinkleff, and A. Wicht, "The concept of white light cavities using atomic phase coherence," Phys. Scr. T. **118**, 85–88 (2005).

[9] R. Fleischhaker, and J. Evers, "Four wave mixing enhanced white-light cavity," Phys. Rev. A **78**(5), 051802 (2007)

[10] H.N. Yum, M.E. Kim, Y.J. Jang and M.S. Shahriar, "Distortion free pulse delay system using a pair of tunable white light cavities," under review, preprint can be viewed at http://arxiv.org/abs/1012.4553.

[11] H.N. Yum, Y.J. Jang, M.E. Kim and M.S. Shahriar, "Pulse delay via tunable white light cavities using fiber optic resonators," will be posted on the arXiv server http://arxiv.org/abs/1012.5482.

[12] In modeling a FP, it is customary to consider each mirror to have an anti-reflection coating on one face, and a partially reflecting surface on the other. However, in reality, the partial reflectivity is produced by using layers of dielectric materials. Thus, in order to model properly any potential phase shift, in reflection or transmission at such an interface, it is necessary to take into account the presence of such a layer, which can be considered to be a Bragg grating. When this is done, it is no longer necessary to consider the presence of another surface with an anti-reflection coating.

[13] H. Kogelnik, "Coupled wave theory for thick holographic gratings," *Bell Syst. Tech. J.,* vol. 48, 2909-2947, (1969).

[14] A. Yariv and P. Yeh, *Optical waves in crystals*, (John Wiley & Sons, New York, NY, 1984)

[15] To calculate the phase shift due to the Bragg reflection, we solved the coupled equations presented in Ref. 13 for the case of $\lambda$(wavelength)=1550nm, $\theta$(angle of incidence)=0, $\phi$(slant grating angle)=0 and $\sigma$(conductivity)=0. The modulated dielectric constant is assumed to be of the form $\varepsilon = \varepsilon_0 + \varepsilon_1 cos\left[(2\pi/\Lambda)z\right]$ where $\varepsilon_0$ is the average dielectric permittivity, $\varepsilon_1$ is the modulation amplitude and $\Lambda$ is the grating period. $\Lambda$ is chosen to fulfill the Bragg matched condition. We have found that the phase shift for the reflected beam is a constant of $-\pi/2$, independent of the vaule of $\varepsilon_1$, which determines the amplitude of the reflectivity. We also found that there is no phase shift for the transmitted wave. Finally, we assumed that the thickness of each BG is infinitesimally small, so that $\Delta\nu_{Bragg} \gg \Delta\nu_{FWHM}$ where $\Delta\nu_{Bragg}$ is a bandwidth of the Bragg reflection and $\Delta\nu_{FWHM}$ is a full width half maximum of the FP etalon. Thus, for a particular $\varepsilon_1$, the Bragg reflection coefficient can be expressed as $exp(-j\pi/2)\sqrt{R}$ with a constant value of *R*, within the spectral range of $\Delta\nu_{FWHM}$.

[16] R.W. Boyd and D.J. Gauthier, "Slow and fast light," in *Progress in Optics*: *Volume 43*, E. Wolf, ed. (Elsevier, Amsterdam, 2002), Chap. 6.

[17] A. Kuzmich et al., "Signal Velocity, Causality, and Quantum Noise in Superluminal Light Pulse Propagation," Phys. Rev. Letts. 86, 18 (2001)

[18] L. J. Wang, A. Kuzmich & A. Dogariu, "Gain-assisted superluminal light Propagation," Nature, 406, 277 (2000)